\documentclass[10pt, osa, twocolumn, nofootinbib]{revtex4-1}

\usepackage{amssymb, amsmath}

\usepackage{lineno}

\usepackage{array}
  \newcolumntype{C}[1]{>{\centering\let\newline\\\arraybackslash\hspace{0pt}}m{#1}}

\usepackage{graphicx}

\usepackage{siunitx}
    \DeclareSIUnit{\molar}{M}
    \DeclareSIUnit{\pixel}{px}
    
\usepackage[pdfborder={0 0 0}, pdftex]{hyperref}

\begin{document}


\title{Polarization of a Bose-Einstein Condensate \\of Photons in a Dye-Filled Microcavity}

\author{S. Greveling}
\affiliation{Debye Institute for Nanomaterials Science $\&$ Center for Extreme Matter and Emergent Phenomena, Utrecht University, Princetonplein 5, 3584 CC Utrecht, The Netherlands}

\author{F. van der Laan}
\affiliation{Debye Institute for Nanomaterials Science $\&$ Center for Extreme Matter and Emergent Phenomena, Utrecht University, Princetonplein 5, 3584 CC Utrecht, The Netherlands}

\author{H. C. Jagers}
\affiliation{Debye Institute for Nanomaterials Science $\&$ Center for Extreme Matter and Emergent Phenomena, Utrecht University, Princetonplein 5, 3584 CC Utrecht, The Netherlands}

\author{D. van Oosten}
\email[Corresponding author: ]{D.vanOosten@uu.nl}
\affiliation{Debye Institute for Nanomaterials Science $\&$ Center for Extreme Matter and Emergent Phenomena, Utrecht University, Princetonplein 5, 3584 CC Utrecht, The Netherlands}

\begin{abstract}
We measure the polarization of a photon gas in a dye-filled microcavity. The polarization is obtained by a single-shot measurement of the Stokes parameters. We find that the polarization of both the thermal cloud and the Bose-Einstein condensate of photons (phBEC) does not differ from shot to shot. In the case of the phBEC, we find that the polarization correlates with the polarization of the pump pulse. The polarization of the thermal cloud is independent of parameters varied in the experiment and is governed by a hidden anisotropy in the system.
\end{abstract}

\date{\today}
\maketitle

\textit{Introduction \label{sec:introduction}} --- In many systems, ranging from condensed matter physics and particle physics to cosmology, phase transitions and spontaneous symmetry breaking play a crucial role. The simplest symmetry to be broken is the U(1) symmetry; the symmetry of the overall complex phase of the wave function of a system. Phases that show the broken symmetry include superconductors~\cite{Luke1998}, superfluids~\cite{Duan1994} and Bose-Einstein condensates (BEC)~\cite{Hall1998, Li2017, Leonard2017}. The associated phase transitions have been observed in many different systems under various conditions, yet the actual breaking of the symmetry can only be observed indirectly, as the absolute phase in quantum mechanics is not a measurable quantity.

Richer physics can be observed when looking at the symmetry breaking of more complex symmetries. In condensed matter physics, beautiful examples have been studied using spin systems in the context of spinor BECs~\cite{Demler2002, Sadler2006}, magnetism~\cite{Kenzelmann2005} and spintronics~\cite{Yao2008, Chen2012}, where symmetry breaking can lead to the formation of magnetic domains separated by domain walls. In these examples, the symmetry is broken because the system chooses a particular direction for the spin degree of freedom.

The realization of BEC of exciton-polaritons~\cite{Kasprzak2006, Snoke2007} and of photons~\cite{Klaers2010, Klaers2011, Nyman2015} opens up new possibilities. After all, both the phase and the direction of the electromagnetic field are observable quantities. In the case of the phBEC, an interesting property of the photon gas is its polarization because the formation of a polarized condensate from an unpolarized thermal cloud constitutes a directly observable example of symmetry breaking. This raises the question: is the polarization symmetry spontaneously broken and hence different for every single phBEC? This subject was recently discussed in theoretical work by Moodie~\textit{et al.}~\cite{Keeling2017}, who developed a model for the polarization dynamics in a dye-filled microcavity which takes into account the polarization states of light, and the effects of angular diffusion of the dye on the polarization state. 

In this letter, we experimentally study the polarization of the photon gas by imaging the Stokes parameters on a single-shot basis. We find that both the thermal cloud and the condensate are polarized. We vary experimental parameters to investigate the breaking of the symmetry. The polarization of the thermal cloud is independent of these parameters, and identical in every shot of the experiment. We conclude it is pinned by a hidden anisotropy in the system. However, the polarization of the condensate is fully determined by the pump polarization. The symmetry breaking is therefore induced.

\bigbreak

\begin{figure}[!b]
  \centering
  \includegraphics[width=0.95\linewidth]{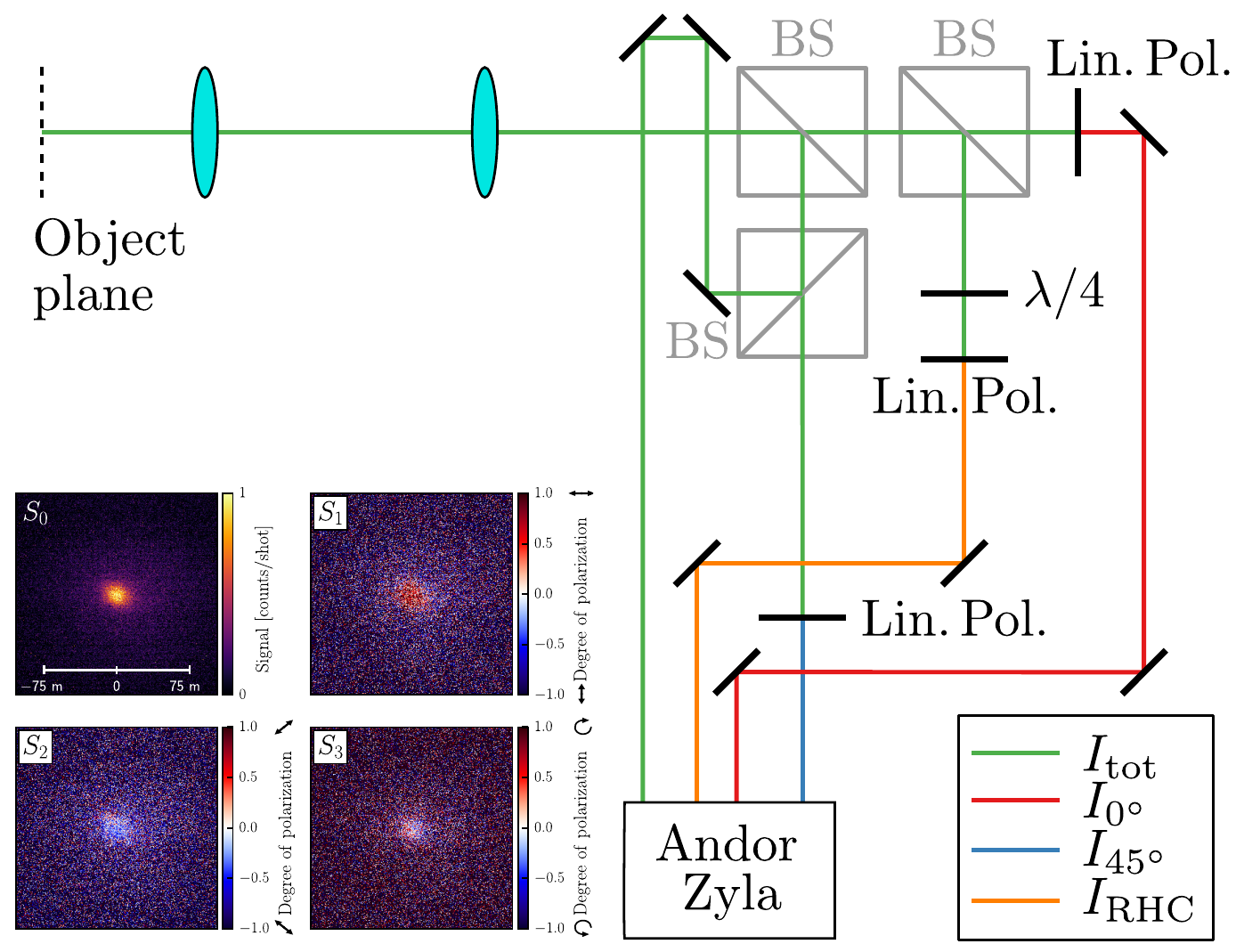}
  \caption{Schematic overview of the polarimetry setup. The colored paths indicate the different polarization states measured on the Andor Zyla. From the intensities we determine the two-dimensional (2D) Stokes parameters. A typical result of the (2D) Stokes parameters, on a single-shot basis is shown in the inset. The contribution to the polarization for each Stokes parameter per pixel is indicated by the corresponding color bar.    \label{fig:Figure_1}}
\end{figure}

\textit{Setup   \label{sec:setup}} --- The polarimetry setup used in this work is shown in Fig.~\ref{fig:Figure_1}. Light escaping through the back mirror of the microcavity enters the polarimetry setup through the object plane, from where it is split and imaged onto four areas of a single camera chip of a scientific complementary metal oxide semiconductor (sCMOS) camera with a dynamic range of 16 bits\footnote{Andor, Zyla 5.5 sCMOS}. The image scale on the camera is \SI{702}{\nano \meter \per \pixel}.
Special care is taken that the four paths from the object plane to the sCMOS chip have the same length, such that all four images are in focus. In each of the paths polarizing elements are placed such that they transmit light corresponding to $I_{\mathrm{tot}}$, $I_{\SI{0}{\degree}}$, $I_{\SI{+45}{\degree}}$, and $I_{\mathrm{RHC}}$, where $I_{\mathrm{tot}}$ denotes the total intensity of the light. The other subscripts denote the angle of linear polarization of light, and right-handed circularly (RHC) polarized light.

With these four intensities, the Stokes parameters are obtained~\cite{Stokes2002}
\begin{equation}    \label{eq:Stokes}
  \begin{aligned}
    S_{0} &= I_{\mathrm{tot}}, \\
    S_{1} &= I_{\SI{0}{\degree}} - I_{\SI{90}{\degree}}, \\
    S_{2} &= I_{\SI{+45}{\degree}} - I_{\SI{-45}{\degree}}, \\
    S_{3} &= I_{\mathrm{LHC}} - I_{\mathrm{RHC}},
  \end{aligned}
\end{equation}
which is one way to fully describe the polarization of light.

The total degree of polarization $p$ is given by the squared sum of $S_{1}$, $S_{2}$, and $S_{3}$:
\begin{equation}    \label{eq:total_degree}
  p = \left. \sqrt{S_{1}^{2} + S_{2}^{2} + S_{3}^{2}} \middle/ S_{0}. \right.
\end{equation}

In order to work with the Stokes parameters, they are combined into a vector known as the Stokes vector $\bf{S}$:
\begin{equation}
  \bf{S} = \begin{pmatrix} S_{0} \\ S_{1} \\ S_{2} \\ S_{3}  \end{pmatrix}\!.
\end{equation}

As one can observe from Fig.~\ref{fig:Figure_1}, the polarimetry setup consists of many optical elements that could effect the polarization of the experimental signal. Since we do not want to measure each individual element separately, we calibrate the setup independently. We define a camera vector \textbf{C} consisting of the signal in the four pixels corresponding to the same physical position. Using the Stokes vector we formally describe the effect of our polarimetry setup as a $4 \times 4$ matrix $\overline{\bf{M}}$ which transforms the Stokes vector into the camera vector,~\textit{i.e.}
\begin{equation}
  \begin{aligned}
    \overline{\bf{M}} \cdot \bf{S} &= \bf{C}, \\
    \begin{pmatrix} m_{00} & \dots & m_{03} \\ \vdots & \ddots & \vdots \\ m_{30} & \dots & m_{33} \end{pmatrix} \cdot \begin{pmatrix} S_{0} \\ S_{1} \\ S_{2} \\ S_{3} \end{pmatrix} &= \begin{pmatrix} C_{\mathrm{tot}} \\ C_{\SI{0}{\degree}} \\ C_{\SI{45}{\degree}} \\ C_{\mathrm{RHC}} \end{pmatrix}\!,
  \end{aligned}
\end{equation}
where $C_{i}$ denotes the components of the vector $\bf{C}$.

Using light sources with known Stokes vectors, we determine the matrix elements $m_{ij}$ for our polarimetry setup. For more details on the calibration matrix, see the supplementary information. After $\overline{\bf{M}}$ is determined, we invert it to transform the camera vectors (for each pixel) into a Stokes vector. We assume that each pixel of the chip is equally sensitive to polarization.

\bigbreak

\textit{Experiment  \label{sec:experiment}} --- During the experiment the photon gas inside the cavity is imaged for each excitation pump pulse of \SI{500}{\nano \second} with a repetition rate of \SI{8}{\Hz}, using the polarimetry setup. For each pump pulse we also take a background measurement. We thus determine for each pump pulse the full polarization of the photon gas inside the cavity. 

The number of photons in the cavity is set by the power of the pump pulse. To investigate whether the photon density influences the degree of polarization, the pump power is increased for each consecutive shot in the sequence keeping the polarization of the pump pulse constant. 

The experimental sequence is performed for different pump pulse polarizations and for a total of three different concentrations of Rhodamine 6G dissolved in ethylene glycol: \num{1.5}, \num{10.5}, and \SI{14.9}{\milli \molar}. As discussed by Moodie~\textit{et al.}~\cite{Keeling2017}, the rotational diffusion constant of the solvent can influence the polarization of the phBEC. We therefore also perform the experimental sequence for Rhodamine 6G dissolved in methanol with a concentration of \SI{1.5}{\milli \molar}. The rotational diffusion constant of methanol is a factor \num{50} lower than that of ethylene glycol~\cite{Siegel2003}.

\bigbreak

\textit{Results \label{sec:results}} --- Using the measured intensities and the inverse of our matrices $\overline{\bf{M}}$,
\begin{figure}[!b]
  \centering
  \includegraphics[width=0.95\linewidth]{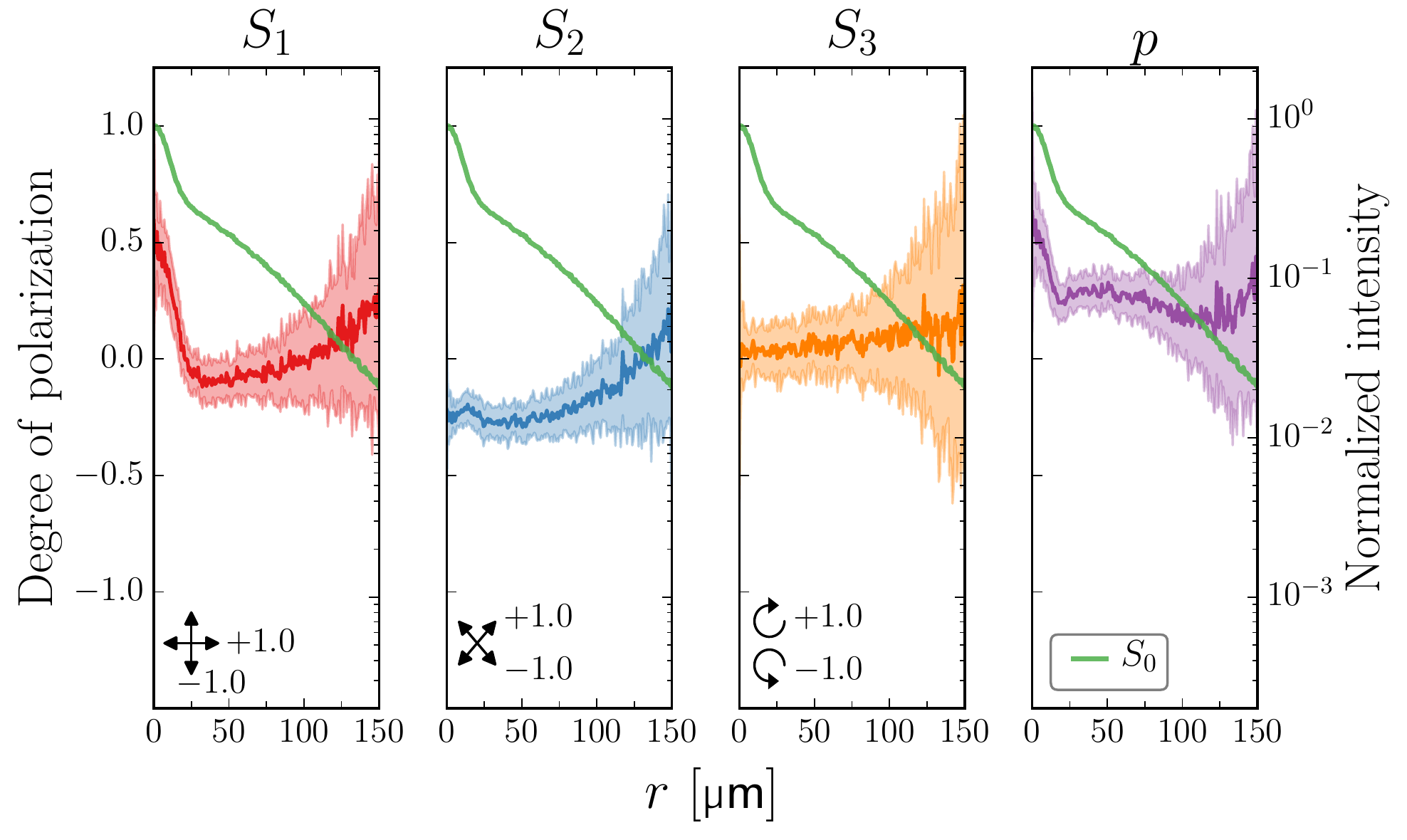}
  \caption{Radial average of 2D Stokes parameters $S_{1}$ (red), $S_{2}$ (blue), and $S_{3}$ (orange) and $p$ (purple), averaged over \num{50} identical phBECs. For each panel the degree of polarization is plotted as a function of the distance from the center $r$. Additionally, each panel also contains the radial average of $S_{0}$ (green), plotted on a logarithmic scale (right axis) as a function of $r$.    \label{fig:Figure_2}}
\end{figure}
we determine the Stokes parameters for every individual pixel. A typical result is shown in the inset of Fig.~\ref{fig:Figure_1}. . Here, the top left image shows the total intensity,~\textit{i.e.} $S_{0}$, normalized to the maximum pixel count. In the false color image, the condensate and thermal cloud are clearly visible. The thermal cloud is identified by the purple color, whereas the phBEC corresponds to the bright yellow center of the image. The other three images in the inset show the three 2D Stokes parameters $S_{1}$, $S_{2}$, and $S_{3}$, which are normalized to $S_{0}$. 

From $S_{1}$, $S_{2}$, and $S_{3}$ one observes a clear difference between the thermal cloud and the condensate, showing that they have different polarizations. In the case of the phBEC, one observes that $S_{2}$ and $S_{3}$ are both close zero, but that $S_{1}$ is close to \num{0.5}, indicating that the phBEC is mostly linearly polarized in the horizontal direction. For the thermal cloud, one observes that all three Stokes parameters fluctuate around zero in the periphery of the images.

Using \num{50} images containing a phBEC, we determine the center of the condensate with subpixel accuracy. From the center we average the data radially outwards for each Stokes parameter. For photon gases created under identical conditions, the 2D Stokes parameters do not differ significantly from one another, which indicates that the polarization of the condensate and the thermal cloud are fixed in our system. We therefore average the radial Stokes parameters over \num{50} BECs created under identical conditions. An example is given in Fig.~\ref{fig:Figure_2}. In the first three panels the Stokes parameters $S_{1}$, $S_{2}$, and $S_{3}$ are plotted as function of the distance $r$ from the center of the trap. In the fourth panel of Fig.~\ref{fig:Figure_2} the total degree of polarization as given by Eq.~\ref{eq:total_degree} is plotted as a function of $r$. The sharp peak close to the center is the phBEC, whereas the exponentially decaying signal for larger distances corresponds to the thermal cloud. The root mean square uncertainty for $S_{1}$, $S_{2}$, $S_{3}$, and $p$ is indicated by their associated pastel color.
\begin{figure}[!b]
  \centering
  \includegraphics[width=0.95\linewidth]{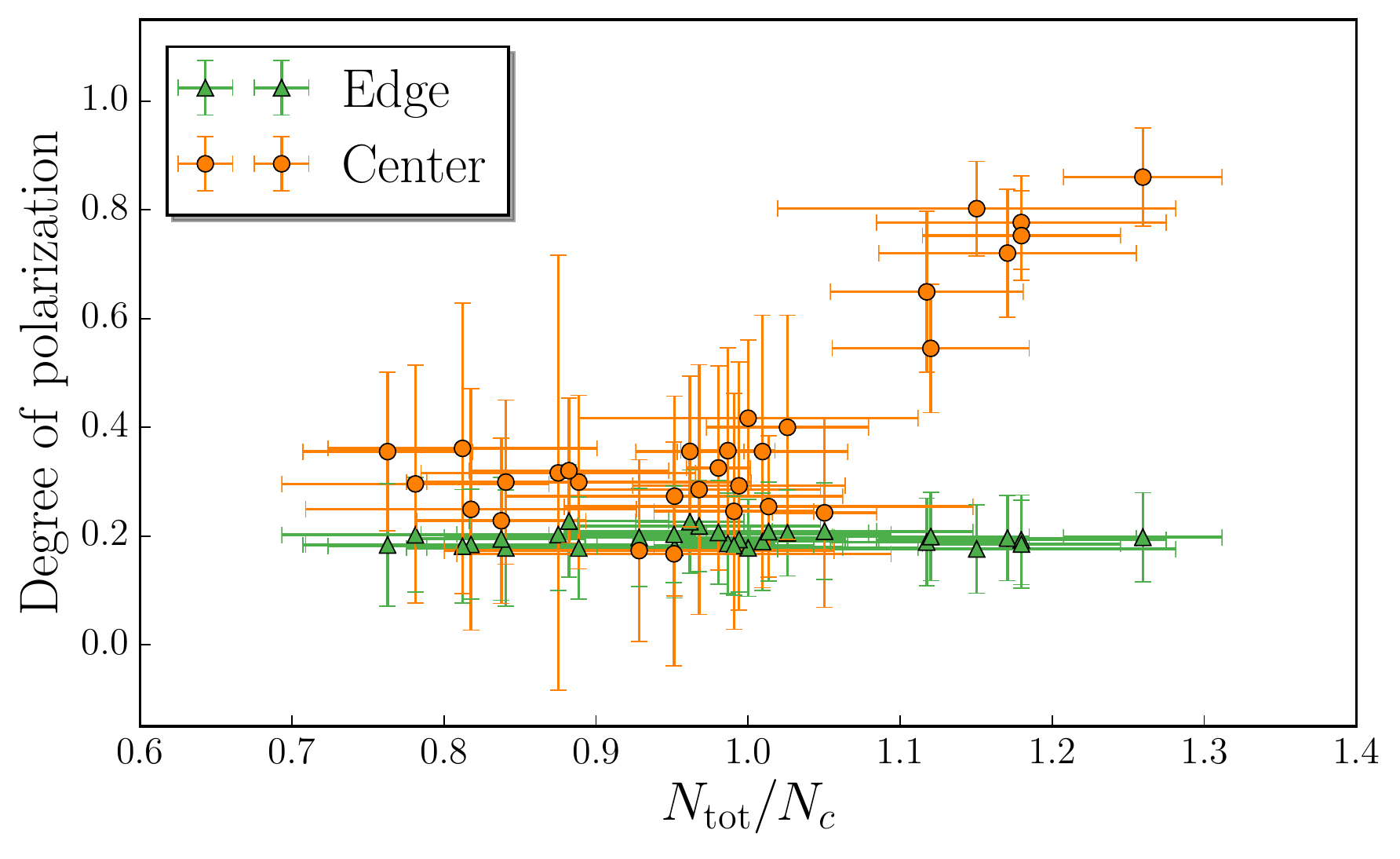}
  \caption{Degree of polarization for the center (orange) and edge(green) of the photon gas as a function of the condensate fraction. When $N_{c}$ is exceeded, a phBEC forms in the center.    \label{fig:Figure_4}}
\end{figure}

From Fig.~\ref{fig:Figure_2} one observes from the fourth panel that the phBEC is strongly polarized with a degree of polarization at the trap center of $p = \num{0.54(23)}$. For increasing $r$ one observes that the degree of polarization decreases. From a range of $r = \num{25} - \SI{115}{\micro \meter}$, the degree of polarization remains approximately constant with a value of $p = \num{0.25(12)}$. As one can observe from the radial profile of $S_{0}$, $r = \SI{25}{\micro \meter}$ corresponds to the spatial size of the phBEC; for larger distances only the thermal cloud remains. For $r \geq \SI{115}{\micro \meter}$, the Stokes parameters and their uncertainty diverge due to near-zero experimental signal in the periphery.

From the first three panels of Fig.~\ref{fig:Figure_2} we observe that the main polarization contribution of the condensate is linear in the horizontal direction. The polarization of the thermal cloud does not have a large single polarization contribution, although $S_{2}$ dominates. 

In Fig.~\ref{fig:Figure_4} we plot the degree of polarization of the center and the edge of the photon gas as a function of the copndensate fraction $N_{\mathrm{tot}} / N_{c}$, where $N_{\mathrm{tot}}$ denotes the total number of photons in the system and $N_{c}$ the critical number of photons. For lower pump powers,~\textit{i.e.} $N_{\mathrm{tot}} / N_{c} < 1$, only a thermal cloud is created. For $N_{\mathrm{tot}} / N_{c} \geq 1$ phBEC is achieved. 

For $N_{\mathrm{tot}} / N_{c} < 1$, the center measurement corresponds to a thermal cloud, but for $N_{\mathrm{tot}} / N_{c} \geq 1$, it also contains a contribution from the condensate, which becomes more important as $N_{\mathrm{tot}} / N_{c}$ becomes larger. The edge measurement denotes the degree of polarization of the thermal cloud, averaged over a range of $r = \num{25} - \SI{115}{\micro \meter}$.
\begin{figure}[!b]
  \centering
  \includegraphics[width=0.95\linewidth]{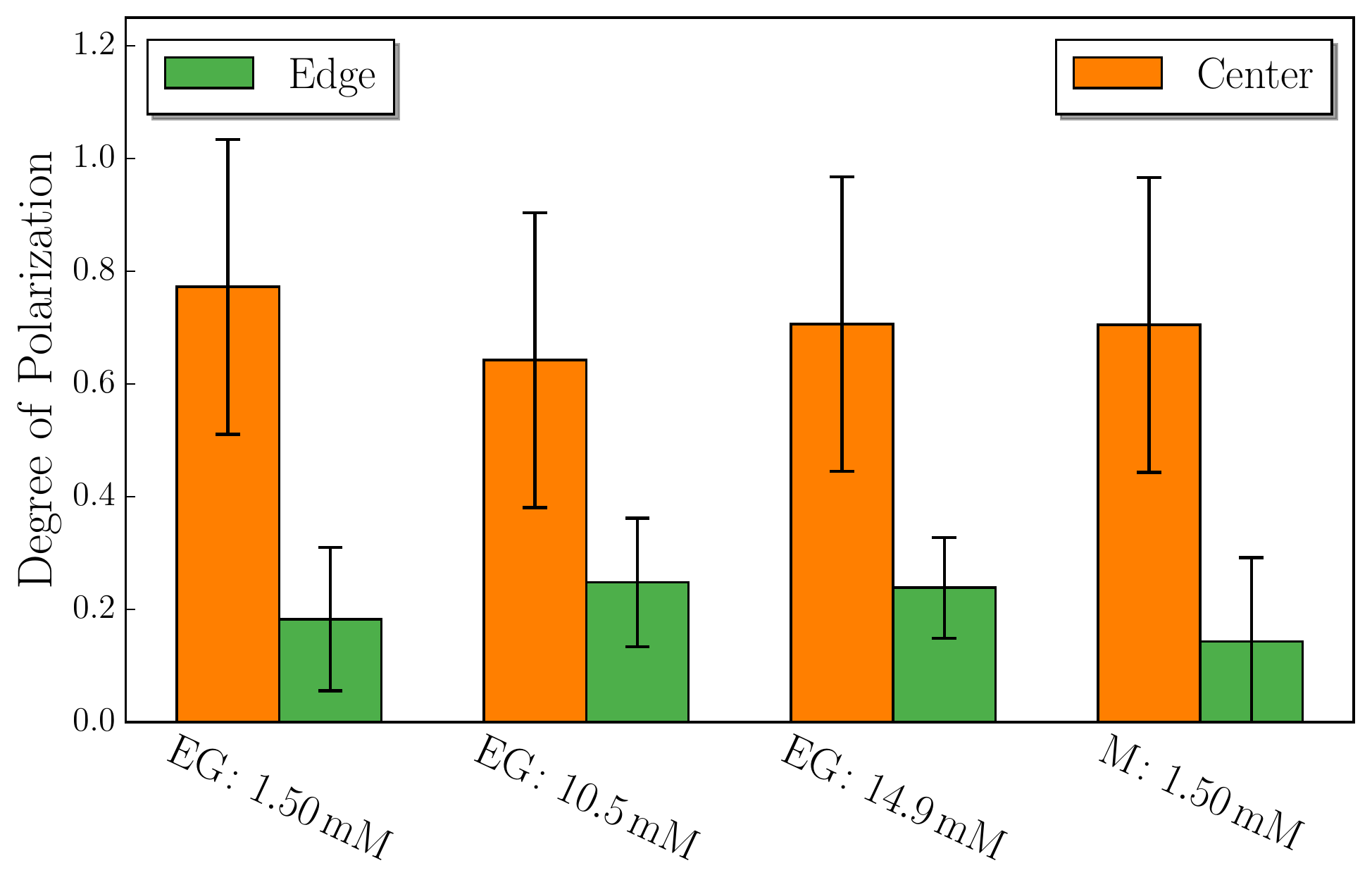}
  \caption{Degree of polarization for the center (orange) and edge (green) of the photon gas as function of different dye solutions and solvents. The average condensate fraction is \num{1.18(7)}.    \label{fig:Figure_7}}
\end{figure}
\begin{figure}[!b]
  \centering
  \includegraphics[width=0.95\linewidth]{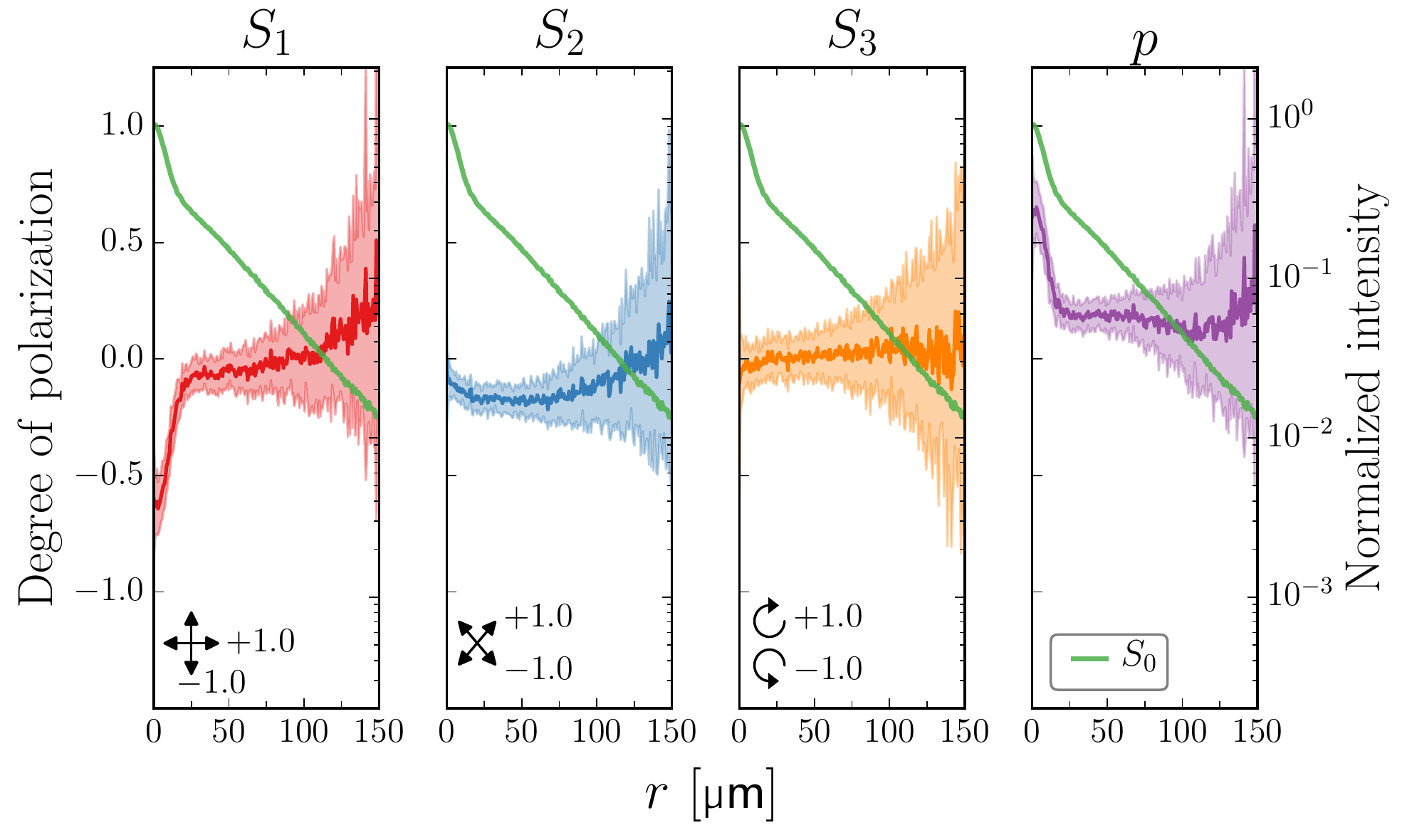}
  \caption{Radial average of 2D Stokes parameters averaged over \num{50} identical phBECs created using a vertically polarized pump pulse. The colors and labels are identical to those of Fig.~\ref{fig:Figure_2}.  \label{fig:Figure_9}}
\end{figure}

For $N_{\mathrm{tot}} < N_{c}$ one observes from Fig.~\ref{fig:Figure_4} that the degree of polarization for the center and the edge of the thermal cloud are the same. For $N_{\mathrm{tot}} > N_{c}$ the degree of polarization in the center differs from that of the edge of the thermal cloud and the degree of polarization in the center increases for increasing $N_{\mathrm{tot}}$. This trend is consistent with the assumption that the condensate is fully polarized in the horizontal direction, as discussed by Moodie~\textit{et al.}~\cite{Keeling2017}.

In Fig.~\ref{fig:Figure_7} we summarize the results of experiments with different dye concentrations and solvents. Each bar in Fig.~\ref{fig:Figure_7} represents the averaged results of phBECs with a condensate fraction of \num{1.18(7)}. As one can observe from Fig.~\ref{fig:Figure_7}, the dye concentration does not significantly influence the polarization results. The degree of polarization and the main contribution remains the same for both the condensate and the thermal cloud for every concentration. Changing the dye solvent from ethylene glycol to methanol also does not significantly influence the results.

All the results above are taken using the same polarization for the pump pulse: horizontal polarization. The radially averaged Stokes parameters obtained using a vertically polarized pump pulse are shown in Fig.~\ref{fig:Figure_9}. Here, the radial profiles are averaged over \num{50} phBECs taken under identical conditions, similar to to Fig.~\ref{fig:Figure_2}.  

From the figure one observes that the phBEC remains strongly polarized with a degree of polarization of \num{0.63(21)}. The result for $S_{1}$ stands out as the polarization contribution described by this Stokes parameter changed from the horizontal direction to the vertical direction. The polarization of the condensate follows the polarization of the pump pulse. The main contributions of $S_{2}$ and $S_{3}$ remain close to zero. The degree of polarization for $r = \num{25} - \SI{115}{\micro \meter}$ remains approximately constant with a value of $p = \num{0.16(12)}$. For the thermal cloud, the contributions have not changed with respect to Fig.~\ref{fig:Figure_2}; the main contribution remains $S_{2}$.

\bigbreak

\textit{Conclusion  \label{sec:conclusion}} --- We investigate the symmetry breaking properties of a phBEC in a dye-filled microcavity by imaging the polarization of the photon gas inside our microcavity, on a single-shot basis. We show that the degree of polarization is identical for every condensate and thermal cloud that we create under identical experimental conditions. For increasing condensate fractions, we show that the degree of polarization at the center of the experimental signal increases. This is consistent with the assumption that the phBEC is fully polarized in the direction of the pump and thus yields a larger contribution for increasing condensate fractions, which is in agreement with the theoretical model by Moodie~\textit{et al.}~\cite{Keeling2017}. The degree of polarization of the thermal cloud is not influenced by varying the condensate fraction, dye concentration, the dye solvent, and in particular is not sensitive to the pump polarization.

We show that the dye concentration or the dye solvent does not influence the results. Changing the pump polarization does not influence the polarization of the thermal cloud. The main contribution remains linear polarization under \SI{-45}{\degree}, independent of the parameters we varied in the experiment. The polarization therefore seems to be governed by a hidden anisotropy. However, changing the pump polarization does change the polarization of the phBEC. The symmetry breaking is thus not spontaneous, but induced by the pump polarization.

\bigbreak

\textit{Acknowledgements} --- It is a pleasure to thank Arjon van Lange, Javier Hernandez Rueda, Erik van der Wurff, Henk Stoof, Peter van der Straten, and Robert Nyman for useful discussions. This work is part of the Netherlands Organization for Scientific Research (NWO).

\newpage
\mbox{ }
\newpage

\section{Supplementary      \label{sec:suppl}}
\subsection{Calibration matrix}
In the case that the optical elements used in our polarimetry setup would be perfect, the calibration matrix would be given by
\begin{equation}
  \begin{aligned}
    \overline{\bf{M}}_{\mathrm{theory}} = \begin{pmatrix} 1 & 0 & 0 & 0 \\ 0.5 & 0.5 & 0 & 0 & \\ 0.5 & 0 & 0.5 & 0 \\ 0.5 & 0 & 0 & 0.5 \\ \end{pmatrix}\!.
  \end{aligned}
\end{equation}

This is however not the case. Using a white light laser and an acousto optic tunable filter (AOTF) we determine the calibration matrix for four different wavelengths; \num{570}, \num{580}, \num{590}, and \SI{600}{\nano \meter}. Using achromatic $\lambda/2$ and $\lambda/4$ waveplates, we determine the Stokes vectors before sending the white light laser through the polarimetry setup. The resulting calibration matrices for each wavelength are
\begin{equation}
  \begin{aligned}
    \overline{\bf{M}}_{\SI{570}{\nano \meter}} = \begin{pmatrix} 0.74 & -0.01 & -0.01 & -0.01 \\ 0.39 & 0.38 & 0.02 & -0.03 & \\ 0.48 & -0.01 & 0.47 & -0.12 \\ 0.39 & -0.01 & 0.33 & 0.22 \\ \end{pmatrix}\!,
  \end{aligned}
\end{equation}
\begin{equation}
  \begin{aligned}
    \overline{\bf{M}}_{\SI{580}{\nano \meter}} = \begin{pmatrix} 0.73 & -0.01 & -0.00 & -0.01 \\ 0.39 & 0.39 & 0.01 & -0.01 & \\ 0.46 & 0.00 & 0.44 & -0.14 \\ 0.38 & -0.02 & 0.32 & 0.19 \\ \end{pmatrix}\!,
  \end{aligned}
\end{equation}
\begin{equation}
  \begin{aligned}
    \overline{\bf{M}}_{\SI{590}{\nano \meter}} = \begin{pmatrix} 0.79 & 0.00 & 0.04 & 0.00 \\ 0.40 & 0.40 & 0.01 & -0.01 & \\ 0.48 & -0.01 & 0.44 & -0.18 \\ 0.40 & -0.02 & 0.36 & 0.18 \\ \end{pmatrix}\!,
  \end{aligned}
\end{equation}
and
\begin{equation}
  \begin{aligned}
    \overline{\bf{M}}_{\SI{600}{\nano \meter}} = \begin{pmatrix} 0.86 & 0.02 & 0.00 & -0.02 \\ 0.44 & 0.44 & 0.01 & 0.00 & \\ 0.52 & -0.02 & 0.48 & -0.20 \\ 0.44 & -0.02 & 0.38 & 0.21 \\ \end{pmatrix}\!.
  \end{aligned}
\end{equation}

The photon gas inside the microcavity consist of a range of wavelengths. Therefore, each matrix is inverted and used to analyze the images obtained with the polarimetry setup. The resulting Stokes parameters are averaged over the four wavelengths.

\vfill

\end{document}